\documentclass[10pt, aas2pp4]{aastex}

\shortauthors{Drake}
\shorttitle{Chromospherically Active Stars}

\begin{document}
\title{Chromospherically Active Stars in the Galactic Bulge}
\author{A.~J. Drake\altaffilmark{1,2}}

%\affil{Dept.~of Astrophysical Sciences, Princeton University, Princeton, NJ 08544}
\altaffiltext{1}{Dept.~of Astrophysical Sciences, Princeton University, Princeton, NJ 08544}
\altaffiltext{2}{Depto.~de Astronom\'ia, P. Universidad Cat\'olica, Casilla 104, Santiago 22, Chile}
%\altaffiltext{3}{Lawrence Livermore National Laboratory, 7000 East Ave, Livermore, CA 94550}

\begin{abstract}
  
  We present the results from the discovery and study of $\sim 3000$ %2962
  chromospherically active giant and subgiant stars toward the Galactic
  bulge. We find that these stars are predominantly RS CVn binaries with
  rotation periods between 10 and 100 days.  We discover that the average
  rotational period of these stars decreases with their distance from the
  Galactic plane.  We find that the primary stars in the RS CVn systems are
  predominantly first-ascent giants.  Our research also suggests that, if
  these stars have spot cycles like the sun, then the cycle period must be
  longer than 10 years on average.  We confirm that the amplitude of the
  spot-induced modulations observed in the light curves of these objects is
  generally larger at minimum light than at maximum.  Furthermore, we
  confirm that the amplitudes of the modulations due to stellar spots
  generally increases as the observed change in average brightness
  increases.  We find no evidence for a relationship between a CA star's
  brightness and its rotational period. However, the average period does
  increase with colour for stars with periods $\lesssim 30$ days.

\end{abstract}
\keywords{ Stars : Spots --- stars : Variables: Other --- Stars: Rotation}

\section{Introduction}

Studies of chromospherically active stars (hereafter CA stars) have been underway
for many years. The two main types of spotted giant stars are single FK Com
stars and binary RS CVn systems (CABs).
FK Com stars are believed to be stars that have coalesced from binaries and 
generally have shorter rotational periods than RS CVn binaries.  Apart from the 
presence of strong Ca$_{\rm II}$ H and K emission both types of CA stars exhibit
photometric variability, soft X-ray and $\rm H\alpha$ emission as well as UV
and IR excess (Hall 1989).  In many cases the secondary stars in RS CVn
systems have orbital periods very close to synchronous with the primary
star's rotation period. However, a significant number are far from synchronous
(Strassmeier, Hall \& Fekel 1993\nocite{Strass93}, Fekel \& Henry 2005\nocite{Fekel05}).  
To date only a couple of hundred of spectroscopically confirmed bright RS CVn binaries 
have been found (Strassmeier et al.~1988\nocite{Strass88}, 1993\nocite{Strass93}).

The advent of large scale microlensing projects such as MACHO (Alcock 1995\nocite{Alcock95},
1997\nocite{Alcock97}) and Ogle (Udalski 1994\nocite{Udalski94}, 2000\nocite{Udalski00}) 
greatly increased scientific knowledge
about forms of stellar variability.  Based on 4 years of photometry of
Baade's Window, \citet{Olech96} found that a large fraction of the {\em Miscellaneous 
Stars} among the OGLE catalog of periodic variable stars
were CA stars.  However, this catalog was comprised of both ellipsoidal
variables and CA stars as a clear distinction between them was not found.
Independently, in on-going studies into the nature of baryonic dark matter
in the MACHO dataset we recently discovered the presence of a number of
spotted variables toward the Galactic bulge and the Magellanic Clouds 
\citep{Drake03}.

A number of detailed studies have been done on bright CA stars.  These
include determinations of the properties of both synchronous and asynchronous
CABs derived from their orbital and photometric periods.  Studies of
their spots have progressed with surface images of the brightest stars. Spot
locations have been derived from high resolution studies of individual
spectral lines \citep{Berd99}.
In general the selection of CA stars can be readily made by using the
Ca$_{\rm II}$ H and K emission strength from spectral surveys.  However, the
photometric properties of CA stars are also significantly distinct from
other forms of stellar variability and they may be discovered from a few
years of photometry.  Clear distinction between RS CVn and FK Com types
requires spectral information.  In this paper we are concerned with the
general properties of CA stars on the Galactic bulge giant branch. We will
use the large numbers of CA stars in this well defined region to examine the
general properties of these stars.

\section{Observations}

The MACHO project repeatedly imaged $\sim$ 67 million stars in a total of
182 observation fields toward the Galactic bulge, the LMC and SMC, to detect
the phenomenon of microlensing \citep{Alcock97}.
These images were taken between 1992 and 2000 on the Mount Stromlo and
Siding Springs Observatories' 1.3M Great Melbourne Telescope with the $8
\times 2048^{2}$ pixel dual-colour wide-field {\em Macho camera}.
In this analysis we will concentrate on the Galactic bulge where $\sim33000$ %32856 
observations of $\sim 33$ million stars were taken in 94 fields.
Each observation field covers an area of $43\arcmin \times 43\arcmin$ on the sky.
Most observations toward the Galactic bulge have exposure times of 150 seconds. 
The observations were taken with a dichroic to give simultaneous images in
MACHO $B_{M}$ and $R_{M}$ bands. These passbands can be converted to the
standard Kron-Cousins system $V$ and $R$ using the photometric calibrations
given in \citet{Alcock99}. The conversion is made to Kron-Cousins $V$ and $R$ bands 
rather than $B$ and $R$, as the transmission of the MACHO $B_{M}$ band is more similar 
to a standard $V$ band filters than a $B$ band ones.
%Alcock et al.~(1999). 
Photometry was carried out using a fixed
position PSF photometry package derivative of the DoPhot package \citep{Schechter93}
called SoDOPHOT \citep{Bennett93}. %(Bennett 1993).

The Galactic bulge was observed from March to October each year.  The seeing of
the data set was generally $> 1\arcsec$ and measurements reach stars in each
field of $V \sim 21$. The photometric sampling frequency varies greatly
between the individual fields with the least sampled fields having only
$\sim 50$ observations while those most sampled contain $\sim 2000$
observations.  In most cases the observations were evenly spaced during the
observing season. Nevertheless, a small number of Galactic bulge fields have
variations in cadence due to a changes in the observing strategy.

\section{Selection}

During our searches for Baryonic dark matter in the form of dark gas clouds
we serendipitously discovered a number of CA stars toward the Galactic bulge
(155) and the Magellanic clouds (12) \citep{Drake03}.
We decided to undertake a specific search to select these variables from the 
MACHO database\footnote{data for variables stars from the MACHO project are 
available at Web http://wwwmacho.mcmaster.ca/.}. To achieve better selection
we first applied a number of cuts to the data to remove noisy or low
signal-to-noise light curves.
The initial step was to select light curves of stars exhibiting a
significant amount of intrinsic variability. This quantity can be defined
as:

\begin{equation}
\hat\sigma = \sqrt{\frac{\sum(\bar{b}_{w} - b_i)^2}{n-1} - \frac{\sum\sigma^2_{b,i}}{n}},
\end{equation}
%Geha et al.~(2004)

\noindent
where $\bar{b}_{w}$ is simply the weighted mean magnitude, and $\sigma_{b,i}$
are the measured uncertainties associated with $n$ individual magnitude 
measurements $b_i$. We chose only light curves where $\hat \sigma > 0.06$ in
both $B_{M}$ and $R_{M}$ passbands. As the stars were observed by the MACHO project 
simultaneously in $B_{M}$ and $R_{M}$, we used the \citet{Welch93}
variability index:

\begin{equation}
I = \sqrt{\frac{1}{n(n-1)}}\sum^{n}_{i=1}\left(\frac{\bar{b}_{w} - b_i}{\sigma_{b,i}}\right)\left(\frac{\bar{r_w} - r_i}{\sigma_{r,i}}\right),
\end{equation}

\noindent
as a second discriminator for selecting the truly-variable objects. This 
statistic ensures that the variability observed in the two bands $b$ and $r$ 
is correlated.  We required a value of $I > 1.5$, as used by \citet{Macho03}
for their QSO candidate selection from MACHO data.

The candidates selected by the first two criteria include many different
kinds of variable stars. In this analysis we are only interested in the CA
stars. Therefore, the other variable star types are considered to be contaminants.  In 
the analysis of \citet{Olech96} 
they noted the difficulty of separating CA stars from ellipsoidal variables 
which occur in the same region of the CMD as CA stars.  Ellipsoidal variables 
exhibit periodic sinusoidal variability with a moderate amplitude similar to CA stars.
However, ellipsoidal variables do not exhibit long timescale changes in
brightness.  In order to remove ellipsoidal variable stars from our
candidate CA stars we used the variogram function of Hughes, Aller, \& Aller 
(1992)\nocite{Hughes92} and \citet{Eyer99}.

The variogram can be determined by comparing pairs of measurements $b_{i}$ and $b_{j}$ 
taken at times $t_{i}$ and $t_{j}$. We first determine $n$ timescale bins over which 
differences in pairs of measurements are calculated by using:

\begin{equation}\label{eq3}
h(n) = n{\rm log}(0.1(t_0 - t_{f})),
\end{equation}

\noindent
where $t_0$ is the date of the first observation for the light curve and
$t_{f}$ is that of the final observation. Within each time bin the median
value, $med[]$, of the difference between measurements squared is determined as:

\begin{equation}
Var(n) = {\rm log}(med[(b_i - b_j)^2])
\end{equation}

\noindent
for bins $h(n-1) < {\rm log}(t_i - t_j) < h(n)$ and $i \neq j$.
We only calculate the differences between measurements for observations 
separated by less than half the total time-span. If the entire time-span was used, 
there would be a strong bias in the median value due to the decreasing number 
of pairs of measurements separated by increasing time differences.
We determine the slope of the variogram:

\begin{equation}
S(n)_{var} = \frac{\delta Var(n)}{\delta h(n)},
\end{equation}

\noindent
for bins $n=3,4,5$, as we are interested in brightness changes on long timescales.
The pairs of measurement in these three bins are typically separated by times longer 
than 100 days. An increasing slope of the variogram selects objects
that are variable on long timescales.
For CA stars, changes in the mean magnitude occur over many years because of the
changes in the sizes and locations of star spots. By comparison
ellipsoidal variables do not exhibit long term variations in baseline magnitude.
For our selection we required that the candidates have a modest increase in 
variability over time, $S_{var} > 0.08$, in each observation band. \citet{Eyer02}
and \citet{Sumi05a}
used $S_{var} > 0.1$ and $S_{var} > 0.09$, respectively, for 
selecting QSO candidates from the light curves of variable objects.
Of the $33$ million Bulge stars monitored the total number of
candidates passing these three variability criteria is $\sim 40000$. %39210.
This set of objects still contains most kinds of long period variable stars
so further selections are required.

%\placefigure{f1}

Observations toward the Galactic bulge contain both disk stars and Bulge stars.
It is difficult to determine the distances of the disk stars photometrically
as they can occur anywhere along the line-of-sight. In contrast Bulge stars
occur in a region $\lesssim 1$ kpc along the line-of-sight \citep{Dwek95}.
In order to select a large set of CA stars at a uniform distance we decided to
choose only CA stars near the Galactic bulge giant branch region. 
In order to remove the disk dwarf stars it is necessary 
to correct the observations for the high reddening observed toward the Galactic
bulge. We used the colour map of Popowski, Cook \& Becker (2003)\nocite{Pop03}
to correct for extinction in each Bulge field. As all the stars that we selected 
were variables we determined the median magnitude for each star and produce a 
CMD from the data. 
In Figure 1 we present the colour-magnitude region selected to
search for CA stars. This region contains  $\sim 11000$ variable stars. %11242
Stars to the left of the selection region are variable stars in the disk while those to
the right are long period variables in the Bulge. The variable stars fainter than this box are a 
combination of faint Galactic disk stars and Bulge stars. The final step in our
selection process is to search for periodic modulations due to spots among the variables 
in this region.

\section{Searching for Periods}

One useful property of CA stars is that their modulations can be used
to determine their rotation periods. Modest changes in the spot induced period 
can occur because of the differential rotation of the stellar surface as well as the 
growth and decay of the spots (Fekel, Henry \& Eaton 2002)\nocite{Fekel02}. 
However, the average periods derived over long timescales should reflect 
the rotation properties of these stars.

For each of the stars in the selection region on Figure 1 we searched for 
periods between 2 and 150 days using the Lomb-Scargle Discrete Fourier Transform 
periodogram method \citep{Scargle82} %(Scargle 1982) 
and Analysis of Variance (AoV) method \citep{Schwarz89}. %(Schwarzenberg-Czerny 1989).
We chose to only look for periods between 2 and 150 days as at shorter periods
the daily observing schedule produces strong aliases while additional aliases 
occur at long periods due to the seasonal nature of the observing. 
Clearly this selection will miss CA stars with the shortest periods while
at long periods we will show there are few such stars.
In addition some of the CA stars are eclipsing while other are not.
The eclipsing CA systems are likely to a different detection efficiency to
non-eclipsing systems.
We treated the $B_{\rm M}$ and $R_{\rm M}$
light curves both individually and in combination.  We found the Lomb-Scargle
periodogram method (hereafter LS method) to be more efficient at finding periods, 
which may be due to the sinusoidal nature of the variability in CA stars.
The LS method also allowed us to determine the probability 
that a period we found was due to a false alarm.
For selection we required that there was $< 1\%$  chance of the period
being due to random noise. However, we note the probabilities derived from
the \citet{Scargle82} method 
were only taken as an approximation. More accurate confidence 
levels can be obtained via Monte-Carlo simulations \citep{Schwarz96}.

%\placefigure{f2}

Light curves that did not exhibit significant periods were removed from
consideration. Without a doubt there are a large number of CA stars among
these as many of the objects removed exhibited long timescale changes in
their brightnesses.  The lack of a significant oscillation
could simply mean that the spots are distributed evenly over the surface of
a star during the observing period. A fraction of CA stars are also likely
to be observed pole-on.  These stars would display little modulation as the
amount of observed spotted area would not change as the stars rotated.
An excessive number of the candidates were found to have periods corresponding to 
the moon's synodic and sidereal periods. The periods of these objects were 
detected because of photometric problems caused by observing fields near the
moon. In most cases these objects were clearly not CA stars 
as they did not have a varying baseline and the periodic variations were not
smooth.  We removed these objects from our selection. 
In addition a large number of period aliases were discovered in field numbers 
104, 119, 301, 402. These were all observed many times per night during the 
last observing season. 
The change in sampling frequency led to a large number of period aliases.  
Therefore, we excluded the data from the final season from the period search. 
As all of these fields have many observations this cut has little effect on 
our analysis. In Figure 2 we present the V and R-band light curve of a
CA star found in the Baade's Window region of the Galactic bulge. We did not
make heliocentric corrections to the times of the observations as these are
have little effect on the results.

\section{General Features of the CA stars}

We examined the light curves of all the candidates by eye and
removed all the light curves that appeared suspicious. During our selection we
also checked accuracy of the periods by phasing the light curves of $\sim 1000$ 
of the light curves. To accomplish this we firstly fitted the light curves
with a lower order cubic spline (polynomial and Fourier fits proved to be
poorly constrained) and subtracted the fits. We then phased the residuals 
with the periods found using the LS method (which was found provide accurate 
results).

%\placefigure{f3}

%Hist.ps
In Figure 3 we present the period distribution of the CA stars.
The longest periods are around 110 days. This fact suggests that even 
slowly rotating stars can have significant chromospheric activity.
There are few CA stars with periods $< 10$ days where FK Com
type variables are expected. This suggest that the bulk of the
CA stars are RS CVn types. In addition, there appears to be some 
evidence of three peaks in the period distribution.

%\placefigure{f4}

After eliminating aliases due to the seasonal observation pattern a number of
the remaining objects still exhibited multiple periods at the $99\%$ level. 
We first investigated the most significant period then the subsequent periods. 
In most cases the additional periods were found to be due to aliases of a 
frequency peaks with very significant spectral power. 
However, it was clear that some of the periods were due to changes in the spots. 
In some cases the periods were twice, or half, the best frequency. Some were clearly 
due to a second group of spots appearing or disappearing at approximately 180 degrees from
the original spot group. This gives rise to two modulations per rotation. In
most cases the longer period was clearly the rotational period when the light
curve was phased. For some light curves we observed that the spot modulations shifted
significantly with time. We believe this is related to the {\em flip-flop} effect 
where regions of activity disappear at one stellar longitude and appear at another
(Berdyugina \& Touminen 1998\nocite{Berd98}; Berdyugina et al.~1999\nocite{Berd99}).  
In Figure 4 we plot the V-band light curve for the object with Macho 
ID 109.19721.291. A smooth low-order spline was fit through this light curve
and subtracted. The residuals have been phased with the 3.2 day period and these 
have been plotted in the lower two panels of this figure. 
The data points in the two lower panels have been separated into the regions 
of different behaviour. The dashed lines in the upper panel mark the limits
of these regions. In the lower left panel we plot the residuals of the points 
marked with dots in the upper panel. In the lower right panel we plot the 
residuals of the points marked by crosses in the upper panel.
Clearly there is a region of transition between the two frequencies at observation
dates between 2400 and 2750 days since JD=2448623.5.
Points in this region are represented by triangles in the light curve but are not
plotted in the lower panels. Clearly there are two real frequencies observed
in this object's light curve. 
This change in the modulation frequency is likely to be due to the presence
of a spot region located $\sim 180\arcdeg$ from a persistent group of spots
(seen at phase 0.75 in the figure). As the main spot group recedes in size,
or increases in temperature, the second group grows in size. This creates
the modulation observed at a phase of $\sim 0.25$. The process appears to
be cyclic as clear repartitions can be seen in the light curve.

\subsection{Eclipsing RS CVns}

As noted above all the light curves were examined to remove uncertain candidate
CA stars. During the examination we found a number of light curves that showed the 
signs of eclipsing binary systems. We expect that any large group of binary stars 
will have some fraction that are eclipsing due to their random inclinations
along our line-of-sight. The shorter the orbital period the more likely an 
eclipse will be visible.  

Measurements of eclipsing systems can be used to determine the orbital period and 
strongly constrain the inclination of the system and the relative sizes of the stars.  
If the reddening and distances are known, as is the case for our bulge giants, 
accurate magnitudes of the stars can be determined and these can be compared with 
spotless stars.  As the spotted regions are eclipsed the times of spot eclipses can 
be used to determine the morphology of the spots.  
If high resolution spectral follow-up is taken, the masses, individual element abundances, 
and spectral types can be determined. This information can then be used to further 
constrain the radii of the stars and therefore the absolute magnitudes and areas covered 
by spots.

The discovery of eclipsing CA stars is not always simple as in most cases the primary stars
are giants of $\sim 10 R_{\sun}$ while the secondaries may be dwarf stars of $\sim 1 R_{\sun}$.
In such cases the change in total brightness due to an eclipse is small. In addition, 
the eclipse is superimposed on a varying brightness level. Eclipses of synchronous 
RS CVn systems are not too difficult to discover as the eclipses occur at a constant 
location in the phased light curves. However, for asynchronous systems the eclipses 
occur at a different relative phase for each stellar rotation. Nevertheless, we did 
discover a number of eclipsing systems simply by examining the light curves. 

%\placefigure{f5}

As an example of the eclipsing systems found, in Figure 5, we present the 
light curve, phased residual curve and phased colour of eclipsing RS CVn system
108.19463.932 (MACHO ID). The RS CVn nature of this object is clear
from the colour, magnitude, light curve modulation and change in mean flux.
We discovered this system has an eclipsing period of 31.2 days. 
The flat bottom of the eclipse shows that secondary star is fully 
eclipsed for approximate 1.5 days each orbit.  This eclipse is accompanied by a 
change in the colour showing that the secondary star is bluer. The depth of the
eclipse is approximately 0.2 magnitudes indicating that it is fainter than the
giant star.  A small secondary eclipse can also be seen in the phased light
curve.  As the phase separation of the eclipses is $\sim0.5$ the orbit is
likely to be circular. The exact properties of the system depend on the
actual masses of the components. However, the properties derived from the light
curves are consistent with a giant of radius $\rm 6.5 R_{\sun}$ and a secondary
with a radius of $\rm 2.0 R_{\sun}$. The rotation period of the primary appears to 
be $\sim 31.6$ days making this a synchronous RS CVn system.

\subsection{Colours, Magnitudes and Locations}

%\placefigure{f6}

We present colours and magnitudes of the CA stars discovered in Figure 6.  
In entire boxed region we found 2962 CA stars from among 11242 variables.
On this figure we also include an empirical estimate of the location of the Coronal 
Dividing Line (hereafter CDL, Haisch 1999\nocite{Haisch99}) with the short-dashed line. 
The CDL marks the division between giants with hot coronae on the left and cool giants 
with massive winds on the right \citep{Henry00}. 
From our sample, 56 CA stars lie to the right of 
the CDL where there are 1638 candidates. There is little likelihood that these 
are not CA stars. Their locations in the CMD are probably due to them lying in regions 
where the amount of extinction is greater than that of the average values given by 
\citet{Pop03}.

%\placetable{tab1}

In Table \ref{tab1} we present the locations, $V$ and $V-R$ magnitudes, 
extinctions, changes in mean brightness $\Delta V$ and $\Delta R$, and 
periods of the CA stars. We also mark the possible and definite eclipsing
CA systems as well as the objects which show signs of the flip-flop phenomenon.

%\placefigure{f7}

The Galactic coordinates of the CA stars are given in Figure 7. The
distribution of CA stars appears to follow the density of Bulge clump giants, 
that is, increasing with decreasing Galactic latitude.  However, at latitudes 
b $> -2.5 \arcdeg$ there are few objects as the amount of extinction is very high. 
In addition, fields having b $< -6 \arcdeg$ contain fewer observations than the
other fields. Therefore, the sample of CA stars is likely to be less complete
than lower latitudes. 

%\placefigure{f8}

%\placefigure{f9}

While investigating the period distribution of CA stars we discovered that there is 
a trend in the period with the Galactic latitude of the stars.  In Figure 8 it 
appears that, on average, the CA stars closer to the Galactic plane have longer periods. 
To better illustrate this we have plotted the period density distribution for
stars in the well sampled region between $-5.4\arcdeg < b < -2.1\arcdeg$ in Figure 9. Here 
we binned the periods in 10 latitude bins by 25 longitude bins. We then smoothed the
distribution with a Gaussian kernel of 0.75 binwidth. This figure shows that, not only is 
there a clear shift in the peak of the period distribution with latitude, but the 
entire period distribution shifts. The slope of the contours make this point clear.
The scale of the period change is approximately 4 days per degree of latitude. It is not 
clear if the relation will hold at large galactic latitudes since this would shift the 
peak of the CA star period distribution to zero.  

The CA star catalog of \citet{Strass93}
gives the locations of $\sim 200$ spectroscopically confirmed binary systems.  
Almost all of these CA systems are within 200 pc of the Galactic plane suggesting that 
the objects in the catalog are predominantly in the thin disk.  At their observed scale 
heights these stars would have $|b| < 2$ at the distance of the Galactic center.
In our experiment it is easier to discover CA stars with short periods near the Galactic 
plane than far from it because these fields were observed more often.
Therefore, if a detection bias exists due to variations in the observing frequency it should 
lead to the detection of more short period CA stars near the disk. The observed period 
distribution shows the opposite effect and is not due to such a bias.

%\placefigure{f10}

In Figure 10 we plot the cumulative period distribution for the 558
CA stars in the region $b > -3\arcdeg$ and 650 stars $b < -5\arcdeg$. We performed a
non-parametric Kolmogorov-Smirnov test on these two cumulative distributions. We 
find the greatest difference between the distributions ($D=0.224$) occurs near
a period of 20 days.  The corresponding probability that these two period distributions 
are drawn from the same population is $P < 9.7\times 10^{-14}$. This is very strong
evidence that there is a change in the period distribution with Galactic latitude.  

%\placefigure{f11}

In Figure 11 we present CMDs of the CA stars in the $b > -3\arcdeg$ and $b < -5\arcdeg$ 
regions. The stars in the high Galactic latitude region are fainter
and bluer than those in the low latitude region.  It is well known that the
average age of Galactic disk stars increases with scale height. However, it is not 
certain whether the Bulge formed rapidly or as the result of several accretion 
processes. 
Based on the analysis of RR Lyrae stars, \citet{Lee92} suggested that the oldest 
stars are in the central bulge region of the Galaxy. In this situation one 
might expect that giant stars at low latitudes would be older and redder 
on average. The two CMDs in Figure 11 do appear to show this effect. 
Additionally, one expects that more evolved stars would rotate more slowly as 
they proceed along the giant branch. The locations of the stars are consistent 
with this possibility. 
However, \citet{Minniti95} also
found a relationship between the metal abundance of Bulge stars and their latitudes. 
This metalicity dependence complicates the interpretation or our result since the 
main-sequence turn-off magnitude is also dependent on stellar abundances.
In addition, most of our CA stars are in binary systems. Such systems evolve differently 
than individual stars because the presence of a binary companion can delay changes in rotational 
velocity as a star proceeds along the giant branch. Contact binary star systems transfer 
mass from the primary star to its companion and this decreases the rotational velocity of the 
primary. However, due to their relatively long periods we expect that most of the CA systems 
are in detached systems. For detached binaries the angular momentum loss of the system can 
cause the orbit to contract and system to spin-up \citep{Maceroni99}. 
As the orbit spins-up the tidal interaction between the stars can cause an 
increase in the rotation rate of the primary. In this picture the more evolved
RS CVn systems could have shorter periods. We note that the increased rotation rate of these
systems can lead to the increased magnetic activity seen in CA stars.
Indeed, orbital periods changes have been observed for many RS CVn systems 
(Hall 1989\nocite{Hall89}; Donati 1999\nocite{Donati99}; Lanza et al.~2001\nocite{Lanza01}). 

\subsection{Brightness and Period Variations}

In order to investigate the general properties of the CA stars we decided
to look at relationships among their common parameters, such as periods, 
dereddened magnitudes, variations in mean brightness, and modulation amplitudes.

The variation in the average magnitude of a CA star, $\Delta M$, is a useful
property since it reflects the change in the degree of spot coverage for a star.
This value can be found by fitting the light curves of the CA stars. However, due 
to the large variations in the number of observations, the periods, and the changes 
in average magnitude, as well as long gaps between observing seasons, we found it 
was not possible to obtain accurate values of $\Delta M$ from automated fits.
Instead, we simply determined the difference between the 95th and the 5th 
percentile of magnitude measurements. These measurement still reflect the variations 
of the light curve while rejecting the brightest and faintest $5\%$ of points.
Discarding the brightest and faintest points removes spurious observations
and minimizes the effect of the sinusoidal modulations superimposed on the
changing mean brightness.

%\placefigure{f12}

In Figure 12 we present the change in light in R-band, $\Delta R$,
versus that in V-band, $\Delta V$.  Clearly the ratio of these values is
very well defined, although there are a small but significant number 
of outliers.
Upon examination of the light curves it was clear that there were cases
where bad data in the R-band observations lead to the outliers 
seen in the figure. Removing the outliers as shown on the figure we find that 
$\Delta R/ \Delta V= 0.900 \pm 0.068$ for 2703 stars, with uncertainty in the mean value 
of 0.0013. %0.00131
This value is consistent with the result for RS CVn giants obtained by \citet{Henry00}.
We note that the inclusion of the outliers does not change the result.
We expect that uncertainties in the transformation from Macho $Bm$ and $Rm$ to 
$V$ and $R$ contribute uncertainty to this relation.  

%\placefigure{f13}

In order to determine whether the amount of activity of the CA
stars changes as they evolve along the giant branch, we decided to
examine the relationship between each star's extinction corrected
V-band magnitudes, $V_0$, and the change in its mean brightnesses, $\Delta V$. 
To remove any biases due to low detection sensitive for fainter 
stars we first excluded stars with apparent magnitudes $V > 18$.
The left panel of Figure 13 shows there is a slight relationship 
such that fainter stars exhibit larger variations in their mean brightness. 
This suggests that faint stars have larger fractional spot coverage than 
brighter stars. There are two possible explanations of this result. Firstly, 
the faint stars have not evolved along the giant branch as far as the brighter stars,
or secondly, these stars are simply fainter due to greater spot coverage. 
The observed average decrease in $\Delta V$ is only $\sim0.06$ magnitudes over 
the range of magnitudes displayed.
To reduce the possible effect of an evolutionary change in magnitude along the giant branch, 
we determined the ridge-line of the giant branch and subtracted this from the observed 
magnitudes. The relationship between this residual brightness, $V_{res}$, and the change 
in baseline magnitude is presented in the right panel of Figure 13.
After the ridge-line is subtracted there still remains some evidence for a relationship between 
the quantities. This result suggests that the trend in $\Delta V$ with $V_{0}$ may simply 
be due to differences in spot coverage, rather than due to evolutionary changes in the 
fraction of spot coverage.

As a separate test we looked at how the properties of CA stars vary with 
rotation period. \citet{Olech96}
divided CA stars from ellipsoidal variables using ``subjective judgment'',
and found that stars with periods $< 10$ days had an upper limit in their I-band magnitude. 
This limit was said to correspond to the limiting ``break-up'' period where a faster
rotating single star would break apart. 
We found no evidence for a limiting V-band magnitude for our CA stars.
We note that long period variables (LPVs) show clear period magnitude relationships 
(Cook et al.~1997\nocite{Cook97}; Wood 1999\nocite{Wood99}, 2000\nocite{Wood00}). 
However, LPVs vary in flux due to pulsation modes not due to surface features. 
We believe that the lack of any such relationship in our analysis is evidence 
that our CA star selection is not contaminated with the ellipsoidal variables
of \citet{Olech96}.

%\placefigure{f14}

In order to further investigate the properties of the CA stars we analysed whether 
there was any relationship between the periods of the CA stars and their colours.
To do this we binned the CA stars as a function of period and determined the mean 
colour of the stars in each period bin. For accuracy the bin size was adjusted
so that there were 50 stars in each bin. In Figure 14 we present the colours 
of the CA stars versus their periods. The averages colours in each bin is shown as 
a triangle along with uncertainty error bars in these values. 
The average colour of CA stars appear to be significantly bluer at periods $< 10$ days 
than $> 20$ days. This may reflect the different nature of the two groups of CA stars 
(single FK Com stars and RS CVns). In the region between 10 and $\sim30$ days the CA 
stars become redder while at longer periods the average colour stays approximately constant.

We also matched the CA stars with objects in the 2MASS all-sky point source
catalog\footnote{http://www.ipac.caltech.edu/2mass/releases/allsky/}. 
We only included matches with measurements in $J$, $H$, and $K_s$ bands and 
found 1077 matches with separations $< 1.5\arcsec$. 
The majority of the matched stars had J-K colours between 0.5 and 0.8 and J band magnitudes 
between 12.5 and 15.  We corrected the near-IR values for extinction using the V-band values
from the \cite{Pop03}'s
map and the conversions given by \citet{Rieke85}.
However, no firm relationship was found between the near-IR colours and the periods of 
the stars.

\subsection{Spots}

Rapidly rotating stars exhibit strong chromospheric activity. Therefore, one might expect
the stars with the shortest periods to have more spots than stars with longer periods.
We investigated this scenario by looking at how the change in magnitude ($\Delta V$)
varies with the rotation period. As noted earlier $\Delta V$ is related to the 
fraction of a star covered by spots. We found that, if there is a relationship
between these quantities, it is below the level of sensitivity obtained with this data.

It has been found that many CA stars exhibit increasing modulations as their
average brightnesses decrease \citep{Henry95}. However, there are also CA
stars that exhibit very small changes in their mean brightness yet show
large changes in their spot-induced modulations (Henry, et
al.~1995\nocite{Henry95}; Fekel, Henry \& Eaton 2002\nocite{Fekel02}; Fekel
\& Henry 2005\nocite{Fekel05}).  The first effect is likely to be caused by
increased spot coverage, since CA stars become fainter as spot coverage
increases and the additional spot area causes the modulation amplitude to
increase.  The second major cause is likely to be due to changes in the
relative positions of the spots with time. The effect of differential
rotation is well known for the Sun. For CA stars this effect could shift the
relative positions of spots occurring at different stellar latitudes. As the
spots line up on one side of a star they produce a change in brightness that
increase with time, yet the average brightness of the star remains
unchanged.  Other minor causes, such as varying spot temperatures and
locations of the spots relative to our line-of-sight, can also cause changes
in the observed brightness.

In order to test how prevalent the effect of increasing spot coverage is on
the modulation amplitude, we investigated the dispersion in each light
curves at maximum light and at minimum light. Although the dispersion of the
data points does not reproduce the actual modulation amplitudes, it does
trace them. The dispersion value is less influenced by outliers caused by
bad data than the maximum and minimum data values.  For simplicity we
assumed that the data points followed a Gaussian distribution and we
determined the associated scatter about the mean measurement value. For each
light curve we included all the data points within a two rotational periods
of the maximum and minimum values. An exception was made for periods $< 15$
days where a 30 day period was used to insure a sufficient number of points
to determine the dispersion. To further improve the accuracy of our result
we removed CA stars with $< 10$ points in this time frame.

%\placefigure{f15}

In Figure 15 we present the difference between the dispersion in
measurements taken near maximum light, $\sigma_{max}$, and those taken near
minimum light, $\sigma_{min}$, for V-band and R-band light curves.  We found
that 1317 of the 2186 (60\%) light curves have $\sigma_{min} - \sigma_{max}
> 0$, where 1093 are expected if there is no relation between the dispersion
and the stellar brightness. Therefore, as the modulation amplitudes directly
follows the dispersion, the amplitude is greater at minimum light with
$4.8\sigma$-level or with $> 99.9\%$ significance.  The same result was also
found for the R-band data. This shows there is little doubt that the large
modulations seen at minimum light are real. Nevertheless, in contrast, this
figure also shows that there are a number of light curves with larger 
dispersions at maximum light than at minimum light.
From our examination of the light curves it was clear that the maximum 
amplitude of modulation does not always occur at minimum light. 
There were light curves with small values of both $\sigma_{min} - \sigma_{max}$ 
and $\sigma_{min}$.

%\placefigure{f16}

To test how the changes in the relative positions of the spots might cause
large changes in the modulations we plotted the dispersion at minimum light
versus the change in average brightness. The result is presented in Figure 16.

There is a strong trend such that the greater the change in mean 
magnitude the larger the average dispersion at minimum light.
Few objects exhibit large dispersions without large changes in 
magnitude. This result suggests that the changes in the modulation
amplitudes are primarily due to the changing spot sizes rather than the 
spots lining up on the surface due to differential rotation.

We attempted to model the evolution of stellar spots on a number of CA stars
using the publicly available spot modeling code, such as {\em SpotModelL} of 
Rib\'rik, Ol\'ah \& Strassmeier (2003)\nocite{Rib03}. We found that this process 
was difficult for our data because the information about the spots is lost between
the end of one observing season and the beginning of the next. The stars with long 
periods may be better modeled in this way as they will have changed less during the 
gap in observing. For the Magellanic cloud the MACHO project made continuous observations
throughout the year. We believe these data would provide much more accurate models
for the range of observed periods.

\subsection{Starspot Cycles}

One aspect of CA stars that has not been conclusively answered is whether they 
have spot cycles similar to the Sun. A number of searches for these star-spot cycles have been
carried out (Berdiugina \& Tuominen 1998\nocite{Berd98}; Lanza et al.~1998\nocite{Lanza98}, 
2001\nocite{Lanza01}; Rib\'rik et al.~2003\nocite{Rib03}).
In some cases cycles have been presented which correspond closely to the time-span 
of the data examined. Cases exhibiting only a single cycle might well be regarded with 
suspicion. Among the CA star light curves we observed, many objects passed through a 
period of maximum light more than once. Whether these truly correspond to a spot cycle 
is unclear to us.

The amount of light observed from a CA star is dependent on the spot
coverage, temperatures, and the locations of the spots.  In general, we
expect that the minimum amount of spot coverage for a CA star will occur
when it is brightest.  The stellar longitudinal distribution of spots
changes the amount of light observed over the period of star.  Therefore,
the mean light and spot coverage can be deduced by averaging the
measurements over the star's rotational period. The movement of spots in
stellar latitude relative to our line-of-sight will also change the observed
light curve, as will variations in spot temperatures. However, these
effects appear less important than the changes in the areas of the spotted
regions.

In order to investigate the evidence for spot cycles and mitigate the secondary effects of 
spot temperature and stellar latitude, we decided to find the average of all the CA star 
light curves. Firstly, we shifted all the light curves in time so that they matched at 
their maximum light. 
Next, we subtracted the maximum light magnitudes from each, and combined
the residual values. The time shifting process essentially gives us twice the time 
baseline of the observations since the time of maximum light is randomly distributed
throughout the observation time.  However, we did not include light curves where maximum 
light occurred within the first or last fifty days of observations, since we do not know 
whether we have actually observed a maximum within the light curve in these cases.  

%\placefigure{f17}

In Figure 17 we present binned residual CA light curve data points
and the average light curve derived from these points for all the CA stars.
As the time of maximum light was chosen to be within our observing season and
there are large gaps between observing seasons, the data points within the 
figure are unevenly distributed.
In this figure each individual data point is the average of points within a
10 day-wide bin. For clarity, only 10 percent of the binned data points are
plotted.  The crosses show the median of the points at each time. These are
representative of what one might expect the archetypal CA star might resemble 
in the absence of spot-induced modulations. The median is presented in
the figure as it better represents the points due to the long tail of faint
values.  We have only displayed the median values that derive from at least
15 data points and 10 different light curves.  However, in most cases more
than a thousand values were used to determine each median value.  

Figure 17 shows little evidence for the median magnitude rising to the
value at maximum light. Although our CA star selection process required an
increasing level of variability up to $\sim 1000$ days, our result
indicates that average CA stars must have a spot cycle periods longer than
2000 days, if they indeed have cycles.
There appears to be some symmetry in the shape of the data point distributions on 
either side of maximum light and median values continue to decrease for $\sim 1500$ days 
on either side of maximum light. The combination of these two findings suggests that 
the average spot cycle should be longer than 3000 days. 
There is some evidence for an upturn in median brightness around 2000 days on either 
side of the maximum light. This may suggest an average spot cycle period of around 
4000 days. However, there are far fewer points at the extremes of this figure, so this may 
simply be a numerical artifact.

Both V and R band (not shown) light curves appear to be in good agreement. The V band 
data shows a larger amplitude as expected from Figure 12. It is possible 
that stars with different rotation rates will have different spot cycle times.
To examine whether there were cycles visible for stars with similar rotation 
periods we investigated the shape of the median curve for subsets of rotation period. 
We found no significant evidence for differences from the combined data presented
in Figure 17.

\section{Discussion}

In this analysis we have found approximately 3000 CA stars.  Without doubt
our catalog does not include all of the CA stars in the observed fields.
From simple geometrical considerations we expect to have found only a few of
the CA stars which are pole-on to our line-of-sight. For such stars a
varying amount of spot coverage, or a change in spot latitude would result in 
a varying baseline magnitude. However, these variations would not create
significant modulations as the CA star rotated. Therefore, these stars will
not exhibit the periodic modulations required for their discovery in this 
analysis. Additionally, as many CA stars
may have many dozens of spots (Eaton, Henry \& Fekel 1996\nocite{Eaton96}),
it is likely that some systems simply did not exhibit strong modulations 
during our observations because their spots were evenly distributed over the 
stellar surface.  Furthermore, many CA stars in fields that were not observed 
frequently will have been missed.

The CA giants we found in this analysis clearly occupy the same region of the CMD as 
clump giant stars. Clump giants are used as standard candles for distances 
\citep{Girardi01}
and as unbiased estimators of microlensing optical depth (Popowski et al.~2005\nocite{Pop05}, 
Sumi et al.~2005b\nocite{Sumi05b}).  The presence of significant numbers of variable stars 
in this region of a CMD might bias the results derived from clump giants. 
We have discovered thousands of variables overlapping the clump region. 
However, \citet{Pop05} found $> 1.3$ million giants observed in these fields,
so the CA variables represent $< 1\%$ of stars in this region. %(Popowski et al.~2005). 
In light of this, we expect that these stars will have little effect on clump distances 
or microlensing optical depths derived from clump stars.
Furthermore, we find no evidence that the CA stars are concentrated in the 
red clump region of the Bulge CMD. %, in agreement with Olech (1996).  
This strongly suggests that they are ascending the giant branch for 
the first time rather than clump giants. As most of our CA stars are located along 
the giant branch, they are mainly within the Bulge rather than the foreground disk.
Our discovery of a relationship between the rotation period and the Galactic 
latitude of CA stars could be due to variations in the population with age and 
metalicity. 

One feature of CA stars is the presence of stellar flares. Flaring events can last up 
to 10 days but are relatively rare \citep{Henry96}.
Because our stars were observed simultaneously in two colours, we decided to search for 
outliers that might be due to stellar flares by looking for spikes occurring simultaneously
in the two light curves. We found no evidence for stellar flares among the thousands of 
light curves. However, we expect that flares are much more likely to be prominent on 
dwarf type CA stars such as BY Dra types.

\citet{Olech96} used low order Fourier fits in an attempt to separate ellipsoidal
variables from CA stars. \citet{Olech96}'s method does not take into account the fact that 
the amplitude of the spot modulations changes with time or the fact that the baseline flux 
varies with time. We found their method to give very poor fits to our CA light curves. We 
believe this is the main reason \citet{Olech96} was unsuccessful at statistically 
separating ellipsoidal variables from CA stars.

\section{Future Work}

We discovered a number of eclipsing RS CVn systems and objects exhibiting
the {\em flip-flop} effect by examining the light curves of the CA stars. 
The Fourier analysis of the light curves would provide a more complete set 
of such objects in our data set.

In order to better understand CA stars it is necessary to study them in a
variety of environments. A study of CA stars in the LMC and the SMC would be
enlightening, as the very limited analysis that was undertaken by
\citet{Drake03} showed some evidence that there are fewer CA stars, per
giant star, in the Magellanic clouds than the Bulge. The reddening toward
the Magellanic clouds is lower and better defined than the Bulge. In
addition, the distances to the Magellanic clouds are well known.
Furthermore, the Macho project continuously monitored stars in the
Magellanic clouds during the term of the project. This coverage would
eliminate period aliases seen in the Bulge due to the seasonal observing
pattern.  It would also allow the maximum and minimum light to be determined
with greater accuracy than in this analysis. The analysis of Magellanic
Cloud data could provide insight into the effects of age and metalicity
variations on CA star periods.

The spectral types of the CA stars discovered here could easily be determined
using a few observations with a wide field multi-object spectrograph such as 
AAOmega\footnote{http://www.aao.gov.au/local/www/aaomega/} on the 4 meter AAT 
telescope. This may provide insights into the nature of CA stars.
The orbital periods of the secondaries could be determined for each of the
CA stars presented. With this information the synchronously rotating RS CVn 
systems could be separated from the asynchronous ones.  In addition, it would be possible
to test whether or not the relationship we found between the period and Galactic latitude 
occurs only for synchronous or asynchronous systems. Synchronous CA systems may show 
trends in the rotational period with age as the strong tidal effects between the stars 
will dictate the observed rotational period. As the primary evolves the redistribution of 
angular momentum is likely to change the orbital and rotational periods. Asynchronous CA star 
systems are more likely to have a rotation periods similar to individual giant stars.
However, this work would require high resolution spectroscopy taken on a number of occasions.

I would like to thank Bohdan Paczy{\'n}ski and Takahiro Sumi for encouraging me to perform 
this work. This paper utilizes public domain data obtained by the MACHO Project, jointly funded 
by the US Department of Energy through the University of California, Lawrence Livermore 
National Laboratory under contract No. W-7405-Eng-48, by the National Science Foundation 
through the Center for Particle Astrophysics of the University of California under 
cooperative agreement AST-8809616, and by the Mount Stromlo and Siding Spring Observatory, 
part of the Australian National University.

%\resizebox{\hsize}{!}{\includegraphics{f1.eps}}
%\includegraphics[width=\columnwidth, height=0.9\columnwidth]{1568f1.eps}
%\centering
%\includegraphics[width=17cm]{f1.eps}

\newpage

%\hspace*{-7cm}
\begin{figure*}
%\centering
\includegraphics[width=2.2\columnwidth]{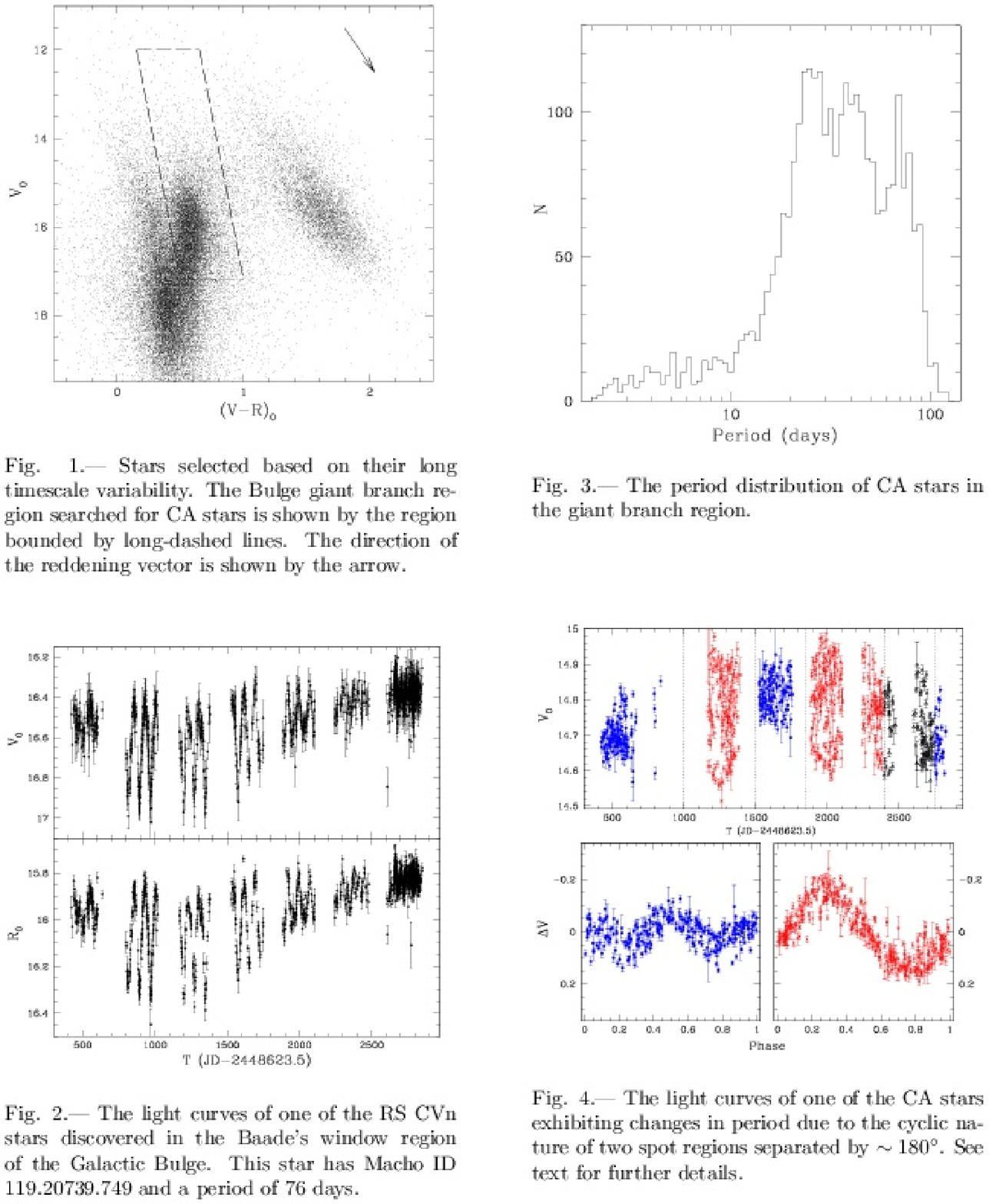}
%\expand 2.0
%\plotone{figs1.eps} 
%\label{f1}
%\footnote{\bf Note the very poor figure quality is due to astro-ph's 1 MB upload limit.}
\caption{\bf Note the very poor figure quality on these figures is due to astro-ph's 1MB submission size limit.}
%\end{figure}
\end{figure*}

\begin{figure*}[ht]
%\expand 2.0
\includegraphics[width=2.2\columnwidth]{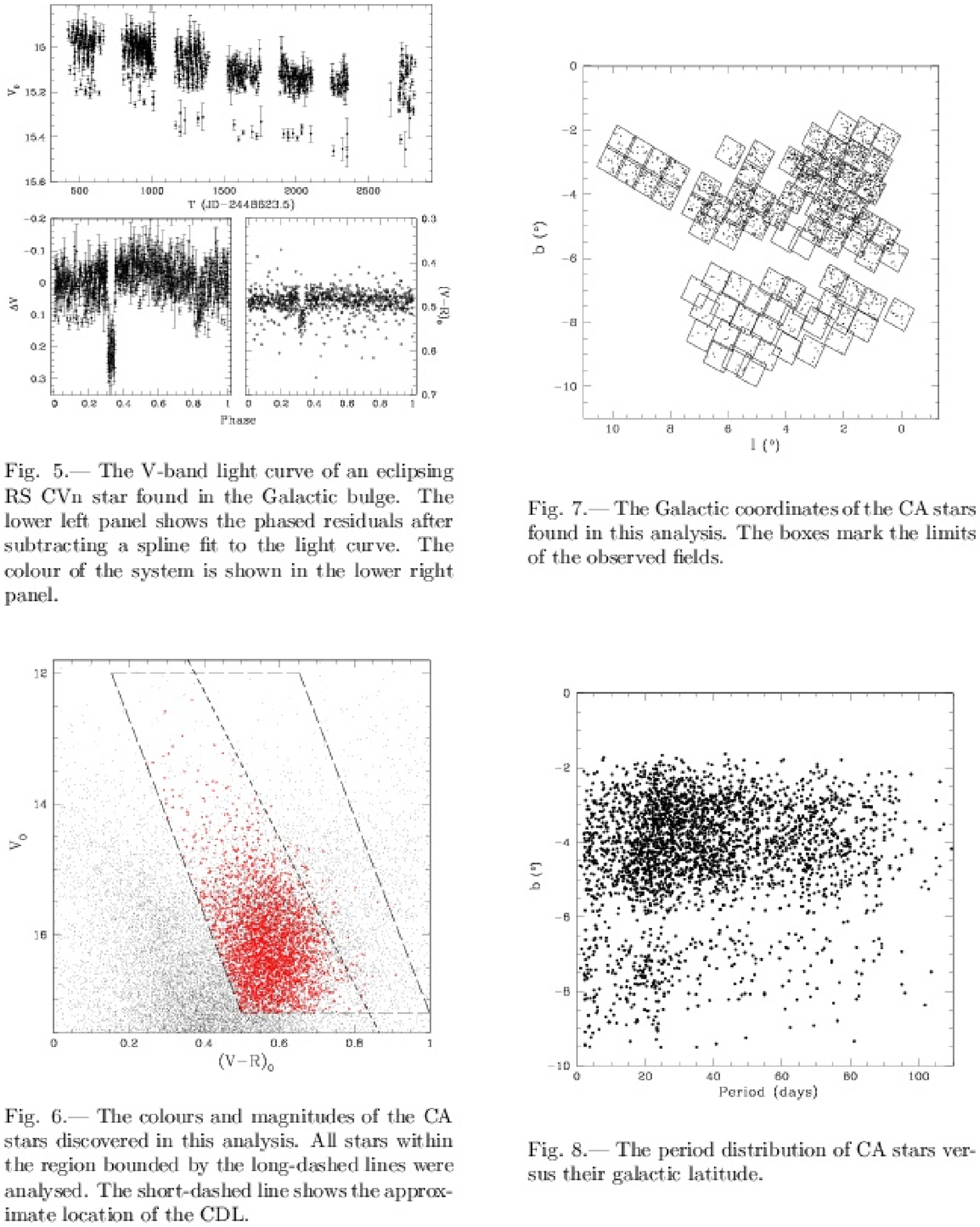}
%\label{f2}
\caption{\bf Note the very poor figure quality on these figures is due to astro-ph's 1MB submission size limit.}
\end{figure*}

\begin{figure*}[ht]
\includegraphics[width=2.2\columnwidth]{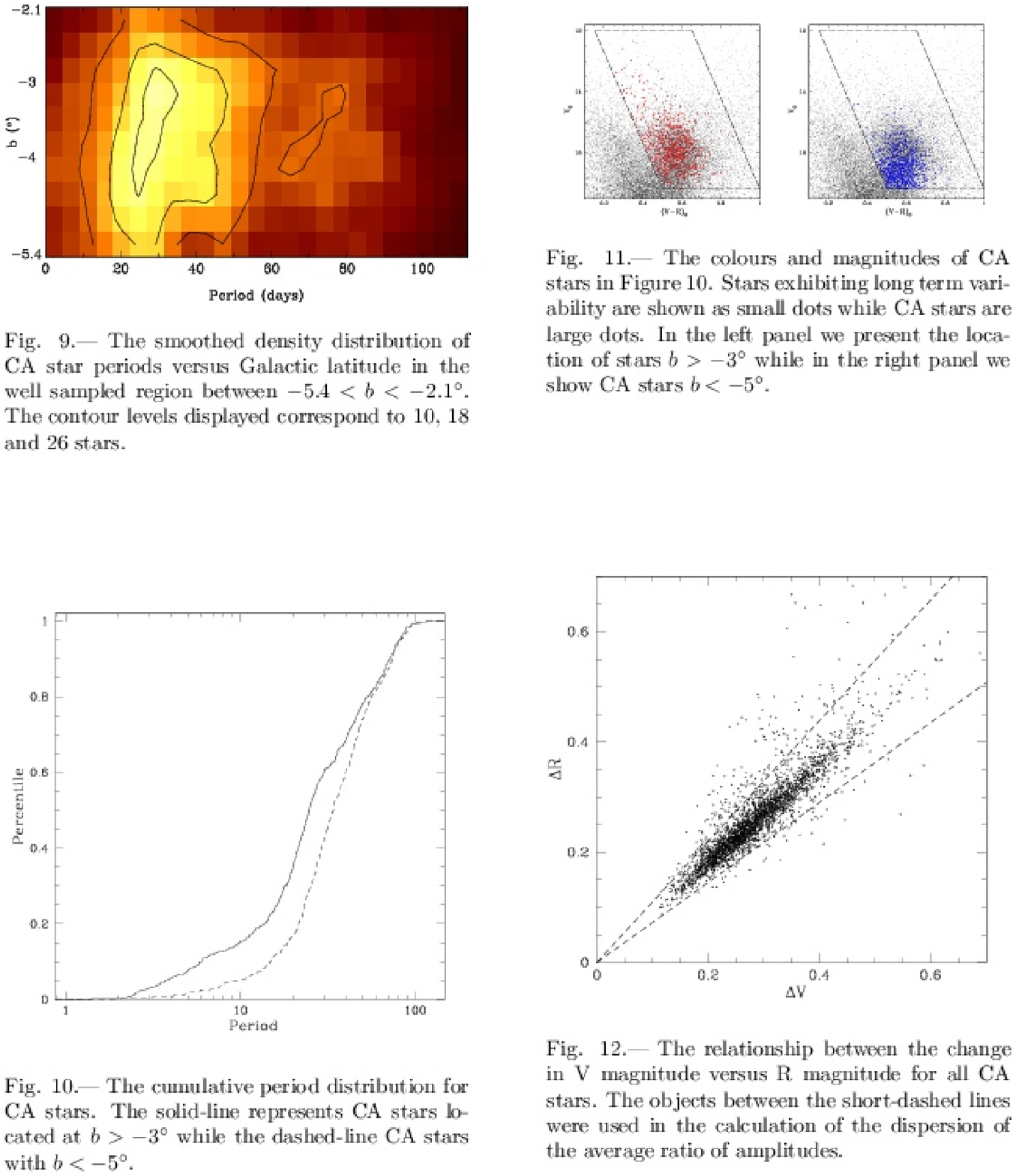}
%  \plotone{figs3.eps} 
%\label{f3}
\caption{\bf Note the very poor figure quality on these figures is due to astro-ph's 1MB submission size limit.}
\end{figure*}

\begin{figure*}[ht]
%\expand 2.0
%\plotone{figs4.eps} 
\includegraphics[width=2.2\columnwidth]{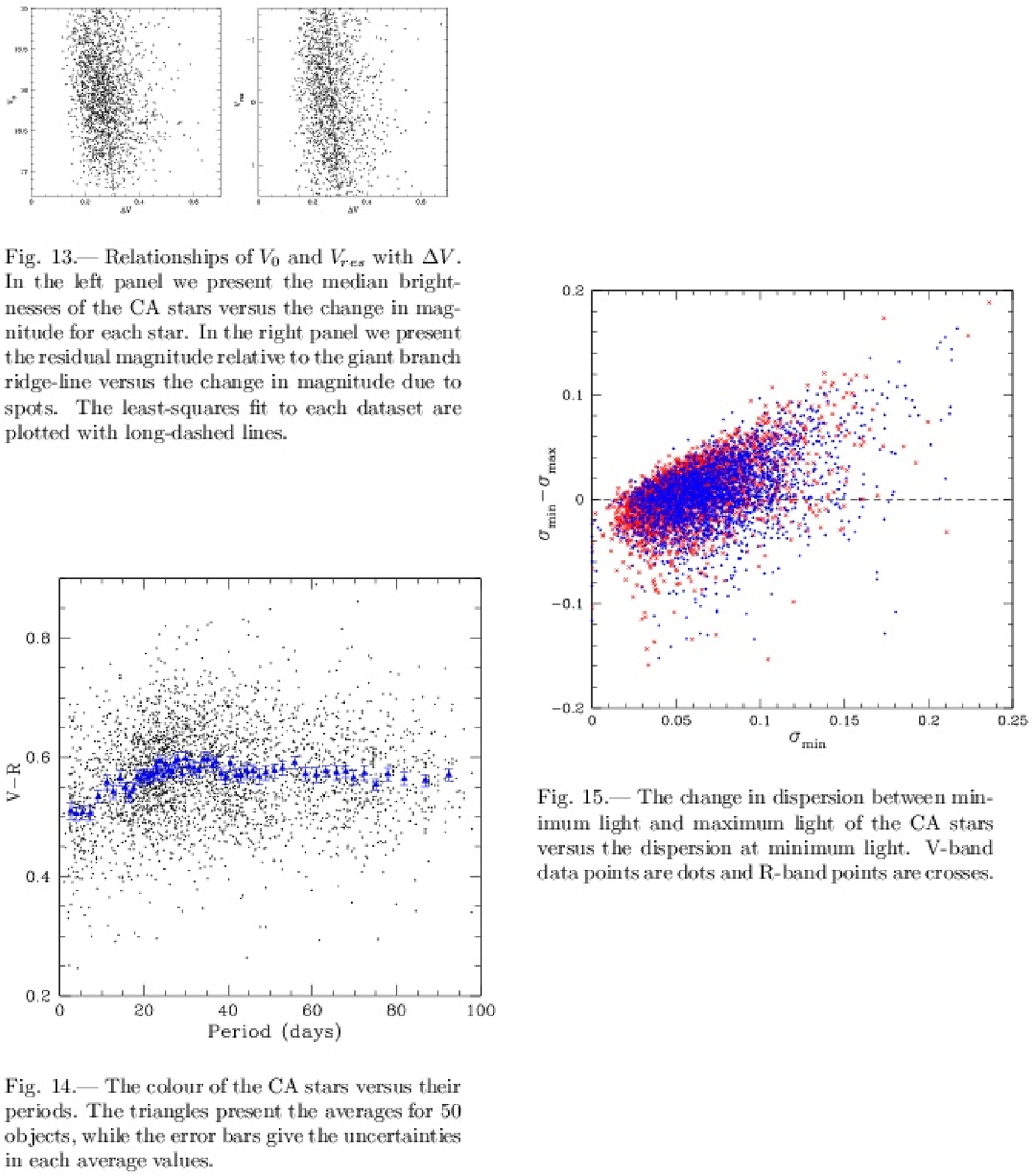}
%\label{f4}
\caption{\bf Note the very poor figure quality on these figures is due to astro-ph's 1MB submission size limit.}
\end{figure*}

\begin{figure*}[ht]
%\expand 2.0
\includegraphics[width=2.2\columnwidth]{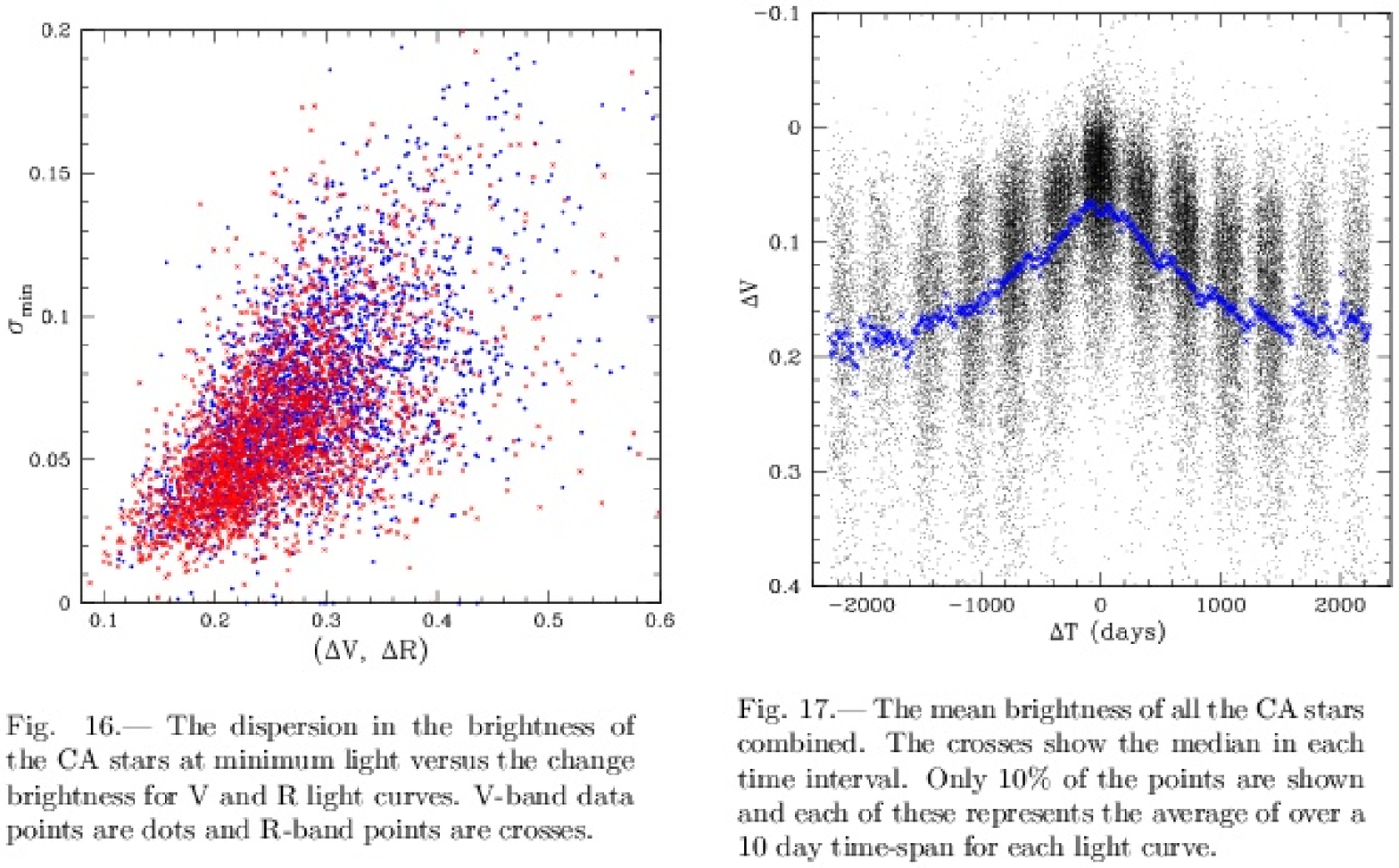}
%\label{f5}
\caption{\bf Note the very poor figure quality on these figures is due to astro-ph's 1MB submission size limit.}
\end{figure*}

%\include{tab1_stub}
    % TABLE1.TEX
\begin{deluxetable}{rrrrccrrr}
\tablecaption{CA Star data.\label{tab1}}
\footnotesize
%\small
\tablewidth{0pt}
\tablehead{\colhead{Macho ID}  &  \colhead{RA} & \colhead{Dec} & \colhead{$V$} & \colhead{$V-R$} & \colhead{$A_{V}$} & \colhead{$\Delta V$} & \colhead{$\Delta R$} & \colhead{P}\\
& h.m.s$\,\,\,\,\,\,\,\,$ & $\arcdeg\, \arcmin\, \arcsec \,\,\,\,\,\,\,\,\,\,$ & & & & & & (days)
}
\startdata
 101.20649.254 & 18:04:26.39 & -27:27:44.9 & 16.95 & 1.11 & 2.18 & 0.47 & 0.41 & 67.465\\ 
  101.20652.878 & 18:04:31.98 & -27:19:38.0 & 17.66 & 0.86 & 1.88 & 0.29 & 0.25 & 30.754\\ 
  101.20653.340 & 18:04:20.86 & -27:12:51.5 & 17.31 & 1.05 & 1.84 & 0.32 & 0.26 & 26.665\\ 
  101.20655.107 & 18:04:25.22 & -27:04:58.4 & 15.99 & 0.89 & 2.04 & 0.22 & 0.20 & 21.803\\ 
  101.20657.810 & 18:04:32.36 & -26:58:44.9 & 18.29 & 1.15 & 2.18 & 0.33 & 0.30 & 49.222\\ 
  101.20778.709 & 18:04:49.72 & -27:35:23.0 & 17.50 & 0.97 & 2.04 & 0.23 & 0.18 & 62.050\\ 
 101.20779.1382 & 18:04:45.18 & -27:30:32.8 & 18.77 & 1.02 & 2.07 & 0.27 & 0.20 & 25.306\\ 
  101.20780.380 & 18:04:48.23 & -27:27:32.7 & 17.18 & 0.98 & 2.07 & 0.28 & 0.21 & 59.574\\ 
  101.20780.731 & 18:04:47.55 & -27:24:37.6 & 17.70 & 0.91 & 2.07 & 0.21 & 0.19 & 33.427\\ 
  101.20780.840 & 18:04:39.84 & -27:24:31.8 & 18.19 & 1.09 & 2.07 & 0.32 & 0.28 & 62.438\\ 
 101.20781.1471 & 18:04:41.27 & -27:22:37.1 & 18.70 & 0.97 & 1.93 & 0.31 & 0.27 & 14.702\\ 
  101.20781.583 & 18:04:36.25 & -27:23:08.5 & 17.37 & 1.00 & 1.93 & 0.29 & 0.27 & 51.504\\ 
  101.20782.371 & 18:04:45.13 & -27:18:13.7 & 16.93 & 1.02 & 1.79 & 0.35 & 0.30 & 93.319\\ 
  101.20783.200 & 18:04:47.67 & -27:12:59.7 & 16.59 & 0.90 & 1.85 & 0.70 & 0.64 & 39.191\\ 
  101.20783.854 & 18:04:39.29 & -27:13:54.5 & 18.19 & 0.96 & 1.85 & 0.36 & 0.30 & 52.013\\ 
\enddata
\tablecomments{Col. (1) gives the MACHO project object IDs. Cols. (2) \& (3) supply the right
  ascensions and declinations at the J2000 epoch.  Cols. (4) \& (5) present
  the median magnitude and colour during observation period.  Col. (6) provides the visual
  extinction derived from the Popowski et al.~(2004)'s colour map.  Cols. (7) \& (8) show
  the maximum change in V band and R band magnitudes during the observing time. Col. (9) states 
  the rotational period of the CA system.
Table 1 will be published in its entirety in the electronic edition of the Astronomical Journal.
The table is also available from author on request.
} %conveys expresses
\tablenotetext{a}{Possible eclipsing binary.}
\tablenotetext{b}{Definite eclipsing binary.}
\tablenotetext{c}{Star which may exhibit the flip-flop effect.}
\end{deluxetable}

\end{document}